%% file: main.tex
\documentclass{vgtc}

\ifpdf
  \pdfoutput=1\relax                   
  \usepackage{graphicx}                
  \DeclareGraphicsExtensions{.pdf,.png,.jpg,.jpeg} 
\else
  \usepackage{graphicx}                
  \DeclareGraphicsExtensions{.eps}     
\fi%

\graphicspath{{figures/}{pictures/}{images/}{./}} 

\usepackage{microtype}                 
\PassOptionsToPackage{warn}{textcomp}  
\usepackage{textcomp}                  
\usepackage{mathptmx}                  
\usepackage{times}                     
\usepackage{cite}                      
\usepackage[hidelinks]{hyperref}
\usepackage{graphics}
\usepackage{comment}
\usepackage{enumitem}
\usepackage{xspace}
\usepackage{color, soul}
\usepackage{amsmath}
\usepackage[title]{appendix}

\def\ie{\textit{i.e.,}\xspace}
\def\etal{\textit{et al.}\xspace}

\def\eg{\textit{e.g.,}\xspace}

\def\vs{\textit{vs.}\xspace}
\def\incl{\textit{incl.}\xspace}

\def\First{\textit{First}\xspace}
\def\Second{\textit{Second}\xspace}

\def\Finally{\textit{Finally}\xspace}
\def\after{\textit{after}\xspace}
\def\before{\textit{before}\xspace}

\newcommand{\prepara}{\vspace{.25em}}

\title{VRContour: Bringing Contour Delineations\\of Medical Structures Into Virtual Reality}


\CCScatlist{
  \CCScatTwelve{Applied computing}{Healthcare information systems}{}{};
}

\author{Chen Chen$^1$\thanks{corresponding author, e-mail: chenchen@ucsd.edu} \quad Matin Yarmand$^1$ \quad Varun Singh$^1$ \\ Michael V. Sherer$^2$ \quad James D. Murphy$^2$ \quad Yang Zhang$^3$ \quad Nadir Weibel$^1$\\ %
{\scriptsize \centering $^1$Computer Science and Engineering and $^2$School of Medicine, University of California San Diego, La Jolla, CA, United States} \\ %
{\scriptsize \centering $^3$Electrical and Computer Engineering, University of California Los Angeles, Los Angeles, CA, United States}}

\input{00-abstract}

\begin{document}
\maketitle

\input{01-introduction}
\input{02-related}
\input{03-preliminary}
\input{04-system}
\input{05-methods}
\input{06-results}

\input{07-discuss}
\input{08-conclusion}

\newpage
\begin{appendices}
\input{appendix}
\end{appendices}

\end{document}

%% file: 00-abstract.tex
\abstract{
Contouring is an indispensable step in Radiotherapy~(RT) treatment planning. However, today's contouring software is constrained to only work with a 2D display, which is less intuitive and requires high task loads. 
Virtual Reality~(VR) has shown great potential in various specialties of healthcare and health sciences education due to the unique advantages of intuitive and natural interactions in immersive spaces.  
VR--based radiation oncology integration has also been advocated as a target healthcare application, allowing providers to directly interact with 3D medical structures.
We present {\it VRContour} and investigate how to effectively bring contouring for radiation oncology into VR.
Through an autobiographical iterative design, we defined three design spaces focused on contouring in VR with the support of a tracked tablet and VR stylus, and investigating dimensionality for information consumption and input~(either 2D or \mbox{2D $+$ 3D}). 
Through a within--subject study~(n = $8$), we found that visualizations of 3D medical structures significantly increase precision, and reduce mental load, frustration, as well as overall contouring effort.
Participants also agreed with the benefits of using such metaphors for learning purposes. 
}

%% file: 01-introduction.tex
\section{introduction}\label{sec::intro}
Radiotherapy~(RT) treatment is indispensable in cancer management.
Contouring is a critical step in the RT treatment planning workflow where oncologists need to identify and outline malignant tumors and/or healthy organs from a stack of medical images (Fig.~\ref{fig::example_interface}).
Inaccurate contouring could lead to systematic errors throughout the entire treatment course, leading to missing the tumors or overtreating the healthy tissues~\cite{Zhai2021}, and could cause increased risks of toxicity, tumor recurrence and even death~\cite{Wuthrick2015}.
Oncology software tools, \eg~Eclipse~\cite{eclipse}, have been introduced to simplify the contouring workflow: oncologists can identify and segment key structures by analyzing and delineating outlines on a stack of 2D scans using a general 2D desktop display (Fig.~\ref{fig::example_interface}b--d). 
This is sometimes not intuitive, and mentally reconstructing 3D physiological structures could introduce significant cognitive load while analyzing 2D images~\cite{He2017}.
%

3D immersive experiences in Virtual Reality~(VR) have unlocked a variety of applications in healthcare~(\eg~\cite{ARTEMIS2021}) and workspaces~(\eg~\cite{Chen2021ExGSense}).
Because of its unique advantages of intuitive and natural interaction, VR has shown great potential in different specialties of healthcare and health science education~\cite{Pelta2022}.
%
%
``Radiation oncology integration'' has also been advocated as a target healthcare application for providers in VR~\cite{Williams2018, Jin2017, Boejen2011}.
%
%
Prior research has evaluated the affordances of VR while being applied to visualize RT treatment plans~\cite{Patel2007, Su2005} and enhance understanding of dose distributions~\cite{Phillips2008}. 
However, the delineation of contours --- the initial phase of the workflow (Fig.~\ref{fig::example_interface}a) --- has not been explored.
%
A few recent works (\eg~\cite{elucis, dicomVR, Chen2022VRcontour}) attempted to bring radiation planning procedures into VR. 
However, the lack of ways to enable precise drawing of contours hinders the practical uses.
While Williams \etal~\cite{Williams2018} showed that $83$\% of oncologist participants found it useful to use VR for visualizing and contouring medical images, the kinds of merits that VR could offer are still unclear.
%

In this work, we present {\it VRContour} and investigate {\it how to bring radiation oncology contouring into head--mounted VR by leveraging the merits introduced by 3D User Interfaces~(UI)}. 
%
%
Since contouring is a challenging and highly specialized task in oncology, we spent more than nine months with domain experts and applied a user--centered design approach to consider their practical needs (\eg~precision and lead time).
%
%
%
Through an iterative and autobiographical design process with healthcare domain experts at different experience levels (\incl~a MD student, a resident and an attending physician), we decided to use a tracked tablet and Logitech VR stylus~\cite{LogitechVRInk} for its high drawing precision, and defined three design spaces by considering the dimensionality (\ie~either 2D or \mbox{2D $+$ 3D}) of {\it information consumption}~(\ie~2D: only visualizing cutting--planes on 2D tablet; \mbox{2D $+$ 3D}: visualizing both cutting--planes on 2D tablet and 3D medical structure) and {\it information input}~ (\ie~2D: only support contouring on 2D tablet with VR stylus; \mbox{2D $+$ 3D}: support contouring on both 2D tablet and direct drawing in 3D).
We populated essential components and metaphors~(\eg~menu designs in immersive environment, 3D object manipulations, and inter--slice interpolations) based on prior works and through our iterative design.

We built a proof--of--concept prototype to evaluate three design spaces on the HTC Vive Eye Pro~\cite{HTCVivePro}.
A separate prototype used to emulate today's contouring on a desktop--based 2D display was also implemented on the same headset as a baseline reference.
To minimize the impact of prior experience with 2D image interpretation, we intentionally involved only junior MD students and other health science students~(\eg~pre--med) who have been trained in anatomy, but without extensive exposure to 2D medical images. 
%
%
Through a within--subject study~\mbox{(n = $8$)}, we showed how 3D visualizations could increase contouring precision, although the capabilities of direct contouring in 3D did not reduce lead time.
%
Our user experience evaluations suggested that VRContour leads to reductions in mental load, frustration, and overall effort when introducing 3D visualizations, and all participants agreed that contouring in 3D is promising for learning purposes.
%
We believe our work will benefit and direct future efforts on designing VR--based oncology software.

%% file: 02-related.tex
\begin{figure*}[t]
    \centering
    \includegraphics[width=\linewidth]{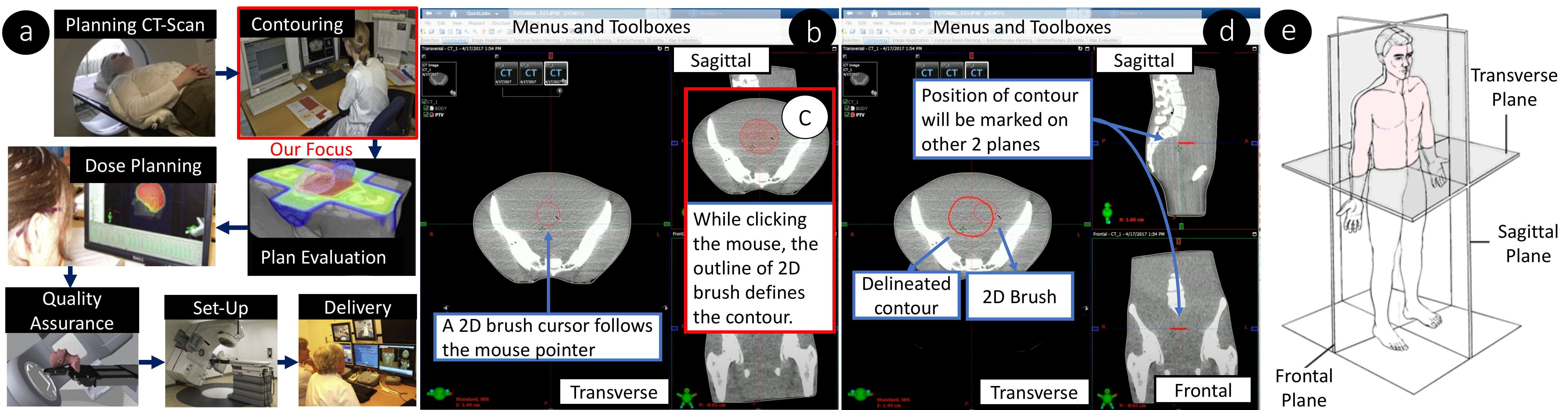}
    \vspace{-0.30in}
    \caption{(a) Workflow of today's RT treatment \cite{Boejen2011}. (b -- d) Contouring using Eclipse~\cite{eclipse}; For delineating contours, a 2D brush is used (b); while clicking the mouse, an area is colored (c); upon releasing the mouse, the contour is defined by the colored area (d). (e) Three planes of motions.}
    \vspace{-0.25in}
    \label{fig::example_interface}
\end{figure*}

\section{Related Work}\label{sec::related}
\subsection{Contouring in Radiotherapy (RT) Treatment Planning}\label{sec::related::contouring}
RT treatment aims to deliver high doses of radiation to a tumor while limiting the dose that is received by the surrounding healthy tissue, which can be categorized into seven steps (Fig.~\ref{fig::example_interface}a)~\cite{Boejen2011}.
Providers need to interpret CT scans~(MRI and PET if needed) to mentally reconstruct a ``virtual anatomy'' and precisely locate and segment tumors as well as surrounding healthy tissues on each slice.
Oncology software, \eg~Eclipse~\cite{eclipse}, is widely used in today's clinics to streamline this process, allowing oncologists to easily align the scans and draw outlines with mouse and/or stylus. 
%
Figures~\ref{fig::example_interface}b--d show an example of the Eclipse~\cite{eclipse} interface, where oncologists can locate and contour the targets by rapidly switching slices across the preferred plane of motion (Figures~\ref{fig::example_interface}e) using scroll wheels.
A 2D brush is commonly used as the delineation tool for drawing and refining contours (Fig.~\ref{fig::example_interface}b--d)~\cite{Werner2012}.
Advanced features, \eg~\mbox{inter--slice interpolations~\cite{Albu2008, Zukic2016nd}} are also used to reduce the number of slices to be manually contoured, leading to less lead time.

Despite these tools, contouring is still a lengthy process, as the oncologists need to examine all slices, each requiring manually distinguishing the tumoral volume from surrounding healthy tissues~\cite{Dowsett1992}. 
%
%
While Artificial Intelligence~(AI) assisted contouring can effectively empower clinicians beyond their current practices~\cite{Zhai2021}, their performance is considered less favourable than manual approaches. 
Additionally, the limited availability of training datasets could impact the scalability and inclusive design of AI--based contouring.
\mbox{Ramkumar~\etal~\cite{Ramkumar2013}} investigated the possibilities of integrating novel interaction modalities into the today's contouring workflow. 
They found that tablet--PC, with multi--touch and gesture--based input are the optimal ones, and could be integrated into the existing workflow for enhancing contouring efficacy.

%
Contouring on today's 2D interface is also error-prone, especially for clinicians with less experience or in lower tier medical centers~\cite{Abrams2012, Duhmke2001, Eisbruch2010}.
%
This is because it requires experience for oncologists to mentally reconstruct the medical structure for target localization by looking at a stack of planar scans, the only support provided by existing tools.
\mbox{Dowsett~\etal~\cite{Dowsett1992}} showed that more than $30$\%~of errors are introduced while contouring spinal canal and lungs using a mouse and lightpen on 2D displays.
%
%
Due to the difficulties of mastering contouring skills, existing research has proposed accessible learning supports using web--based \cite{eContour,Sherer2019} and mobile applications~\cite{Yarmand2022astro}.
While enhancing contouring training might be helpful, we believe that the new immersive 3D visualization technology can potentially help oncologists simplify and streamline the contouring procedure, and therefore might lower the barriers and efforts of contouring procedures in the long term.
%

\subsection{Precise Drawing in VR}\label{sec::related::vr_drawing}
%
Enabling oncologists to engage in precise VR delineations is challenging.
%
While most existing works focus on drawing for 3D modeling, many design components are transferable to our context.

Early studies have investigated the issues of drawing inaccuracies in 3D~\cite{Schmidt2009}.
\mbox{Fitzmaurice~\etal~\cite{Fitzmaurice1999}} indicated how orientations of pen and paper could affect the 2D drawing performance.
\mbox{Arora~\etal~\cite{Arora2017}} showed how the lack of drawing surface support is the major cause of inaccuracies, which also depends on the orientations of drawing surface.
\mbox{Cannav\`{o}~\etal~\cite{Cannavo2020}} and \mbox{Chen~\etal~\cite{Chen2022}} demonstrated the merits of using stylus for 2D and 3D drawing respectively in VR while being compared to traditional VR controller.
%

%

%
With the aforementioned VR based drawing evaluations, several existing works proposed drawing tools by integrating 2D tablets for general designers.
For example, SymbiosisSketch~\cite{Arora2018} proposed a 3D drawing application by combining the interaction metaphors of mid--air drawing and surface drawing, as as to create detailed 3D designs with head--mounted AR.
They showcased how designers could create a 3D canvas with mid--air stroke, and subsequently refine and add details with 2D tablet.
%
%
%
%

While focusing on designing contouring applications for oncologists, our work borrowed several design ideas from these works.
%
However, due to the uniqueness of contouring requirements and the needs of practicing oncologists, directly transferring the techniques from applications facing general users to contouring applications does not necessarily lead to a usable system by radiation oncologists.
We therefore co--designed with domain experts iteratively to better understand our system requirements~(Sec.~\ref{sec::prelim}).

\begin{figure*}[t]
    \centering
    \includegraphics[width=\linewidth]{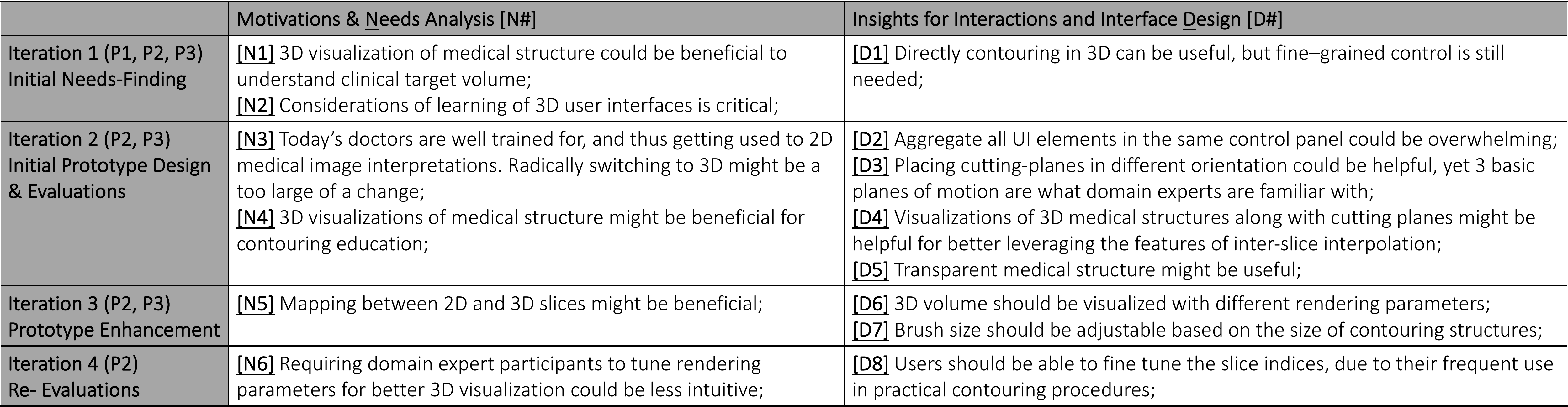}
    \vspace{-0.25in}
    \caption{Insights and takeaways from preliminary interviews and iterative design.}
    \vspace{-0.25in}
    \label{fig:iterative_design_takeaway} 
\end{figure*}

\subsection{XR in Healthcare and Health Science Educations}\label{sec::related::vr_health}
%
%
%
Extended reality~(XR) has been advocated as the next frontier for future healthcare and health science education~\cite{Balasubramanian2021}.
Existing works have explored the affordances of immersive 3D visualization of medical structures and their real--world applications.
Zhang~\etal~\cite{Zhang2021} built a server--aided 3D DICOM viewer using mobile Augmented Reality~(AR), with real-time rendering, interactions and full transfer functions editing.
However the lack of contouring support limits its use for RT treatment planning.
Anatomy Studio~\cite{Zorzal2019} proposed an AR tool for virtual dissection that combines tablets with stylus to assist anatomists by easing manual tracing and exploring cryosection images.
While supporting manual segmentation, the limited resolution and the non--transparent isosurface rendering prevents anatomists from exploring finer grained structures.

As for health sciences education, Nicholson~\etal~\cite{Nicholson2006} have demonstrated significant merits of using VR for anatomy education, with an average of $18$\%~of score improvement.
Ma~\etal~\cite{Ma2016} designed an AR applications by overlaying the CT dataset on top of real--world participants, and demonstrated $86.1$\% approval in terms of educational value.
%
Boejen~\etal~\cite{Boejen2011, Su2005} summarized how VR could benefit today's RT training by enabling students to simulate and train clinical situations without interfering with practical clinical workflow and without the risks of making errors. 

In terms of RT treatment planning, Patel~\etal~\cite{Patel2007} showed how the stereoscopic visualization could foster the understanding of \mbox{spatial} relationships between anatomy and calculated dose distributions.
Williams~\etal~\cite{Williams2018} suggested that 3D visualization could be advantageous when the definition of the clinical target volume is based on a 3D spread of disease along anatomical features with intricate topology.
Such volumes could be difficult to observe and evaluate as they usually do not lie neatly in any of the motion of planes~\cite{Williams2018}. 
DICOM VR~\cite{dicomVR, Williams2018} showed a system that allows oncologists to manually segment the target directly in 3D space.
Early results of using this system show that while the contouring time seems to be reduced compared to 2D display based software, the drawing precision might be sacrificed to some extent~\cite{Williams2018}.
Although $83$\% of oncologists found DICOM VR useful~\cite{Williams2018}, the underlying reasons are still obscure. 
%
%
In addition, the lack of physical surface support and use of VR controllers naturally hinders precise delineations while refining outlines on a 2D virtual plane~\cite{Arora2017}.
We aim to understand how VR can benefit contouring, and how to design such a system by fully exploiting the affordances that this medium offers.

%% file: 03-preliminary.tex
\begin{figure*}[t]
    \centering
    \includegraphics[width=\linewidth]{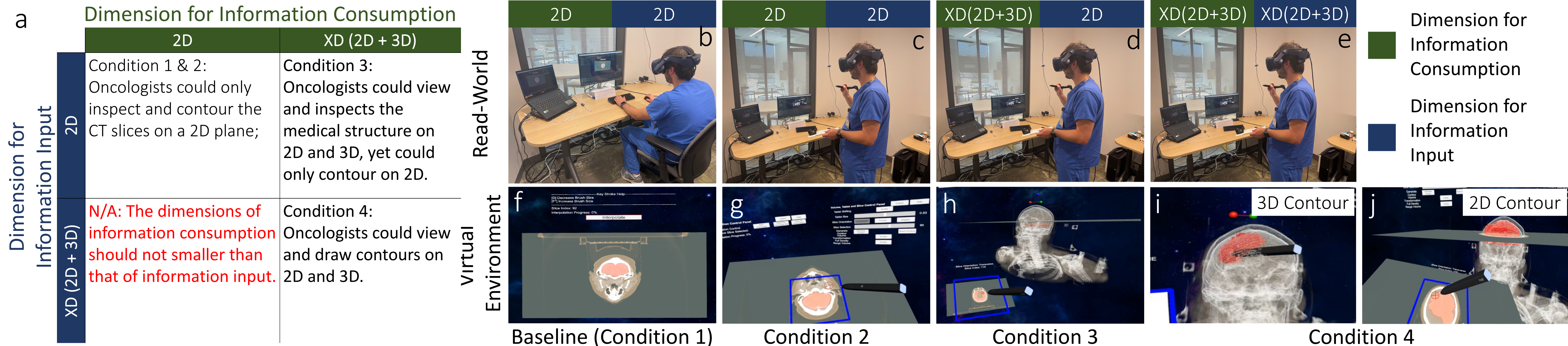}
    \vspace{-0.30in}
    \caption{Design of VRContour. (a) Explorations of the design space by considering the dimension of information input and consumption. (b -- j) Final prototype of four conditions generated by our preliminary autobiographical and iterative design process.}
    \vspace{-0.25in}
    \label{fig::design-space-explorations}
\end{figure*}

\section{Preliminary Exploratory and Iterative Design}\label{sec::prelim}
%
%
We used an autobiographical approach and user--centered design, to explore and then implement as well as refine our prototype {\it iteratively} with domain experts~\cite{Neustaedter2012DIS, Chen2022VRcontour}. 
%
%
All studies (\incl~Sec.~\ref{sec::method}) have been approved by the Institutional Review Boards~(IRB).

\subsection{Participants}\label{sec::prelim::participants}
Our participants include a third year medical student (P1, F), a senior resident fellow (P2, M), and an attending oncologist (P3, M).
P2 and P3 have extensive experience teaching contouring, and all participants are affiliated to the \href{http://health.ucsd.edu}{UC San Diego Health System}.
No participant had prior VR experience.
%
As today's contouring on 2D display is highly dependent on familiarity and understanding of 2D radiology images, the results from P2 and P3 are highly affected by their past experience regarding interpretations of medical scans and 2D contouring workflow.
To minimize such impacts we therefore included P1 intentionally, who is familiar with human anatomy, but not with 2D medical image interpretations.

\subsection{Procedure}\label{sec::prelim::procedure}
Our preliminary design consisted of four iterations. 
Each interview (or co--design workshop) lasted for $30$--$90$~minutes (\mbox{$M$ = $53$~min.}, \mbox{$SD$ = $30$~min.}).
%
%
Thematic analysis was used for data interpretation.

\prepara
\noindent{\bf [I1] Iteration 1 --} 
We first conducted three remote semi--structured interviews with all participants~\cite{Yarmand2021}.
We explained basic VR concepts by showing example VR based medical image visualization applications, to help them brainstorm conceptually. 
We focused on two guiding questions: \mbox{\textbf{(i)}} {\it ``What are the benefits that VR might bring to today's contouring?''} and \mbox{\textbf{(ii)}} {\it ``What are the features that might be useful to augment during contouring procedure?''}

\prepara
\noindent{\bf [I2] Iteration 2 --}
We then built a simple prototype based on \mbox{[I1]}, where a tracked tablet and Logitech VR stylus was used as the input tool (Fig.~\ref{fig::tablet_sizing}). 
Participants could draw contours on a cutting--plane rendered on a tracked tablet using a 2D brush metaphor~(Fig.~\ref{fig::drawing}a -- d), which is consistent with today's 2D contouring~(\mbox{Fig.~\ref{fig::example_interface}b -- d}). 
We also designed a 3D brush metaphor by extending the existing 2D brush, which is essentially a semi--transparent sphere attached to the tip of the VR pen.
Participants could then directly draw a colored volume inside 3D medical structures using a 3D brush~(Fig.~\ref{fig::drawing}e--f).
All control menus were attached to the tracked tablet.
Section~\ref{sec::design} outlines the details of our design.
P2 and P3 were invited to a co--design workshop where they could try out the system; P1 was excluded from this second iteration due to the lack of familiarity with practical contouring procedures.

\prepara
\noindent{\bf [I3] Iteration 3 --}
We improved the system design based on lessons learnt from \mbox{[I2]}, and organized remote interviews with P2 and P3. 
The goal of the session was to understand the feasibility and setbacks of the provisional design.

\prepara
\noindent{\bf [I4] Iteration 4 --}
We finally invited P2 to experience the enhanced system.
We also discussed extensively with P2 to ensure domain experts participants could inspect the target structure in both 2D and 3D through our rendering algorithm. 
P3 was not involved during [I4] due to clinical duties.

\subsection{Findings}
Fig.~\ref{fig:iterative_design_takeaway} shows key takeaways during each iteration.
\textbf{[N\#]} and \textbf{[D\#]} denote {\it the motivation \& \underline{N}eeds analysis}, reflecting experts' thoughts of how 3D contouring might be beneficial to future oncology diagnosis and educations, as well as suggestions on {\it interaction and interface \underline{D}esign}.
%
%
%
%
The lessons learnt are summarized into three areas:

\prepara
\noindent \textit{\textbf{Inspecting Medical Images in 3D --}}
All participants recognized the challenges of learning how to analyze 2D medical images, especially for small organs ({\it``things like vasculature can be harder to identify.''}~[P1]).
Participants explained the approach to locate and understand a specific target from a set of medical images: 
{\it `` You get a stack [of planar medical images]. And then you can go [back and forth], and change the different [cutting plane] to see where you are. [...] We don't get to see only one image. You can go up and down and then kind of orient yourself in.''}~[P1].
Key benefits include the ``overall view'' capabilities for helping determining the boundary between tumors and surrounding healthy tissues ---{\it``[...] often you're trying to decide where does that cancer stop? [...] Being able to visualize everything in 3D could help you see those borders better.''}~[P2]---and having flexibility for placing and orienting cutting--planes for inspecting target as well as inter--slice interpolations [P2, P3].

%

\prepara
\noindent\textit{\textbf{Annotating \& Contouring Medical Images in 3D --}}
All participants, during I2 and I4, agreed that directly drawing in 3D could be helpful for identifying hard tissues (\eg~bones), yet not for the soft ones (\eg~blood vessels).
P3 emphasized the importance of preserving the features of 2D contouring, yet using 3D as an assistive visualization technique:
{\it ``as you're drawing in 2D, you can see a 3D rendering of a structure [...] that actually could be a really interesting concept and might improve contouring accuracy. To really see the structure, like a prostate cancer, being able to see the course of the rectum, see where the prostate sits, see how the seminal vesicles go over [...] and kind of understand that anatomy, and actually see that 3D, that actually might result in more accurate contouring, especially for people that are just learning.''}
P2 further extrapolated the usefulness of mapping 3D annotations back to 2D slices:
{\it [It would be nice to] have like a bright star, and you can put it somewhere. And then when you switch over [to 2D] and start scrolling up and down, [you see] a star on the excellent position where you put it. } [P2]

\prepara
\noindent\textit{\textbf{Learning about 3D UI and Interactions --}}
During [I1], P2 and P3 took on average $10$~minutes to get used to the prototype.
They believed such learning effort should \textit{not} be problematic for general practicing oncologists (\eg~{\it ``I don't think it would take that long to get used to pushing the buttons''}~[P2]).
Participants also emphasized the importance of leveraging the skills of inspecting and contouring medical slices on 2D displays that oncologists already have in today's practice: {\it``we work with 2D images, but have to think in 3D. It's a little bit of paradox.''}~[P3], {\it``No doctor will ever see it as 3D, we're always scrolling a plane [...] but it's always on 2D.''}~[P2].

\subsection{Discussions and Design Space Explorations}\label{sec::prelim::design_space}
%
%
We finally defined a design space, leading to three evaluation conditions $+$ a baseline condition for emulating today's 2D contouring~(Fig.~\ref{fig::design-space-explorations}a).
The prototypes of the four conditions are demonstrated in Fig.~\ref{fig::design-space-explorations}b--j.
We re--use the concept of information flow as the dimensions of our morphological analysis~\cite{Sears2009HCI}.
Specifically, the {\it dimension of information consumption} indicate the ways that the users consume the output information from VR (\eg~inspecting scans), and the {\it dimension of information input} refers to how users outlines contours inside VR (Fig.~\ref{fig::design-space-explorations}a).
While each dimension can be either 2D or \mbox{2D $+$ 3D}, the dimension for information consumption should not be lower than the one of the information input.
Based on \mbox{[N3]} and \mbox{[D1]}, we also intentionally excluded any design that would enable contouring {\it purely} on 3D, as mid--air drawing in VR makes it challenging to be precise due to the lack of physical support and perception of depth~\cite{Arora2017}.
The four conditions are summarized below:

\prepara
\noindent{\bf [C1] Contour on a 2D desktop display (Baseline) --}
We built a prototype to emulate today's contouring on 2D desktop display using a keyboard and mouse~(Fig.~\ref{fig::design-space-explorations}b+f). 
Since enhancing the comfort of the VR headset is beyond our scope, we re--created this scenario to establish a baseline, such that the confounding factors introduced by comfort of VR headset could be minimized.
While wearing the VR headset, oncologists need to go through each transverse slices to mentally understand the medical structures, followed by outlining key structures on each slices.
During \mbox{[I4]}, we confirmed with P2 that the prototype could sufficiently emulate today's 2D contouring.

\prepara
\noindent{\bf [C2] Contour on a 2D tablet {\it \textbf{without}} visualizations of 3D medical structure --}
C2 leverages the idea of direct interaction where the drawn stroke is aligned with the physical tip of the VR pen~(Fig.~\ref{fig::design-space-explorations}c+g). 
Participants are required to inspect and contour on 2D slices rendered on a tracked tablet, using a VR stylus.
No 3D visualizations were provided in this condition.

\prepara
\noindent{\bf [C3] Contour on a 2D tablet {\it \textbf{with}} visualizations of 3D medical structure --}
C3 offers a 3D rendered medical structure hovering in the mid--air (Fig.~\ref{fig::design-space-explorations}d+h). 
This, hypothetically, could reduce the cognitive task load for thinking in 3D while inspecting different cutting--planes in 2D. 
As oncologists are not allowed to draw in 3D, we consider the dimensions for information input as 2D only.

\prepara
\noindent{\bf [C4] Contour in the 3D medical structure {\it \textbf{and}} on tablet --}
Participants are expected to contour on both 3D and 2D (Fig.~\ref{fig::design-space-explorations}i+j). 
Besides the hypothesized benefits in C3, we extrapolated that the direct 3D contour could reduce lead time, due the potential elimination of slice--wise drawing on the 2D interface.
After 3D contouring, oncologists could continue refine the outlines with pen and tablet.

%% file: 04-system.tex
\setlength{\abovedisplayskip}{0pt} 
\setlength{\abovedisplayshortskip}{0pt}

\section{System Design}\label{sec::design}

\begin{figure*}[t]
    \centering
    \vspace{-0.15in}
    \includegraphics[width=\linewidth]{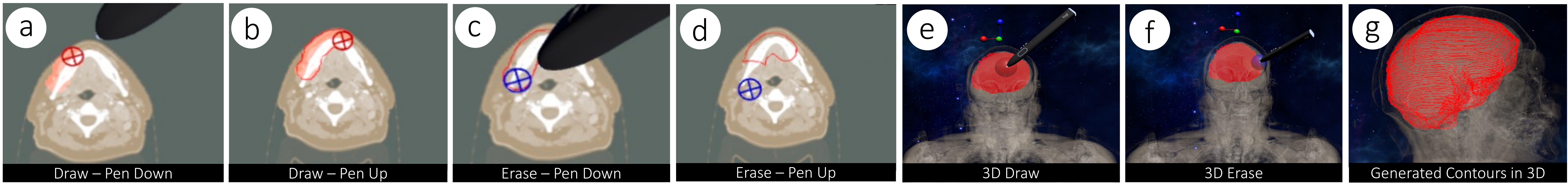}
    \vspace{-0.25in}
    \caption{Drawing/erasing contours using brush metaphors. (a-- b) The contour is automatically generated after the VR pen is lifted from the tablet; (c -- d) the new contour is re--generated upon the completion of erasing; (e -- f) contours, visualized as a volume, could also be drawn/erased directly inside 3D volume; (g) generated contours in 3D (side view).}
    \vspace{-0.25in}
    \label{fig::drawing}
\end{figure*}

\subsection{Volumetric Rendering}\label{sec::sys::rendering}
%
%
Although investigating optimal rendering techniques is beyond our scope, existing works (\eg~\cite{Zhang2021}) have shown that there is no single algorithm that could be applied to every 3D task (also confirmed in \mbox{[D4]}).
While considering the needs of transparent rendering \mbox{[D5]}, we decided on using a ray--casting approach that treats the 3D medical structure as a transparent object that could transmit, emit and absorb light~(like a cloud).
Although Direct Volume Rendering~(DVR) and Maximum Intensity Projection~(MIP) are commonly used approach to render transparent volumes, MIP has a well known limitation regarding depth perception~(\eg~ambiguous front and back relations)~\cite{Ney1990, Mady2020}.
We thus decided on using DVR.
As we focus on understanding the merits that the 3D UI could offer to contouring, and the tuning of rendering parameters is important \mbox{[D6]} yet beyond our scope, we therefore pre--adjusted the rendering parameters for each evaluation task before final evaluation (see \mbox{Step 1} below and Fig.~\ref{fig::rendering}). 
This process was done during \mbox{[I4]} with P2.
The implementations of rendering pipelines include three steps:

%
\prepara
\noindent{\bf Step 1: Data Prepossessing \& Sampling --}
We represent the 3D medical structure using a 3D array, by stacking all \mbox{DICOM} slices.
To emulate practical contouring, we retain the original resolution of each slices without either upsampling or downsampling.
The physical space between voxels over each dimensions are determined by the scanning machine.
%
%
We used min--max normalization to scale the density at each voxels to $[0, 1]$ and the resultant view of 3D visualizations are called {\it ``full density range view''}.
During \mbox{[I4]}, we adjusted such range (\ie~vary the lower and upper density bound) to make sure that the target structures are visible (\ie~{\it ``fine density range view''}, see also Fig.~\ref{fig::rendering}).
The ray, cast from camera, would then sample the volume into $256$~equal--distance steps that were chosen based on visual effects and system performance.

\prepara
\noindent{\bf Step 2: Classification --}
Classification refers to the step of using transfer functions to assign each discrete step with color and opacity.
%
For simplicity, we only consider the simplest 1D transfer function to map the normalized (or adjusted) density to a RGBA vector. 
%
%
%

\prepara\noindent
{\bf Step 3: Composition --}
Finally, we aim to generate a composited value by blending the RGBA at each step along the ray, which would be eventually rendered on the 2D projection.
We used the \textit{over} operator~\cite{Porter1984} to compute the final RGBA vector.
The algorithm for front-to-back traversal could be referred to equations~(\ref{eq::c}--\ref{eq::alpha}), where $\mathbf{c}_i$ and $\mathbf{\alpha}_i$ refer to the color and opacity at $i^{th}$ step ($i \in \{1, 2, 3 ... 256\}$).

\vspace{-0.1in}
\begin{equation}
    \mathbf{c}_i = \mathbf{c}_{i - 1} + \mathbf{c}_i\mathbf{\alpha}_i(1 - \mathbf{\alpha}_i)
    \label{eq::c}
\end{equation}
\vspace{-0.15in}

\vspace{-0.15in}
\begin{equation}
    \mathbf{\alpha}_i = \mathbf{\alpha}_{i - 1} + \mathbf{\alpha}_i(1 - \mathbf{\alpha}_i)
    \label{eq::alpha}
\end{equation}
\vspace{-0.15in}

More details in terms of volumetric rendering techniques are outlined by Lichtenbelt \etal~\cite{Lichtenbelt1998}.

\subsection{Interaction Techniques}\label{sec::system::interaction}

\begin{figure}[b!]
    \centering
    \vspace{-0.10in}
    \includegraphics[width=\linewidth]{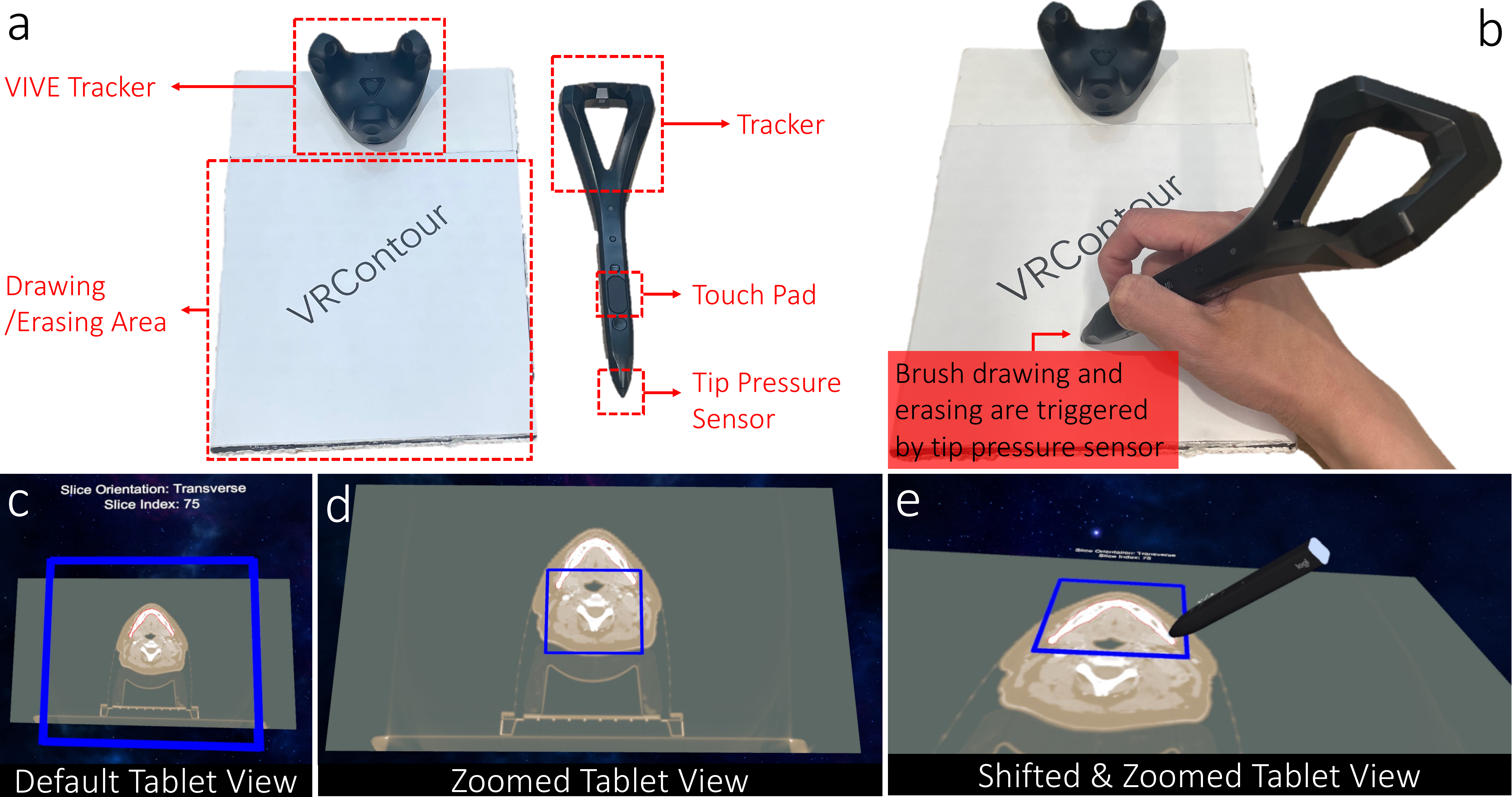}
    \vspace{-0.25in}
    \caption{Contour on 2D. (a) Tracked tablet and VR pen; (b) real--world scene using VR pen and a  tablet;  (c) a blue frame indicating the physical boundary of the tablet; (d) the cutting planes could be zoomed in for gaining a detailed view; (e) oncologists can only draw inside the blue frame due to the physical surface support.}
    \label{fig::tablet_sizing}
\end{figure}

\noindent{\bf Pen + Tablet Interactions and Supporting Precise Contouring --} 
We decided to use the \href{https://www.logitech.com/en-us/promo/vr-ink.html}{Logitech VR stylus}~\cite{LogitechVRInk} as input tool in both 3D and 2D contouring for its high precision, which meets the need for precision drawing in a virtual environment~\cite{Arora2017}~(Fig.~\ref{fig::tablet_sizing}a).
We used a tracked physical board, with same size as a $10.2''$~iPad, since having a physical support should increase the drawing precision~\cite{Arora2017}~(Fig.~\ref{fig::tablet_sizing}a--b). 
Participants could scale the size of the virtual tablet to inspect the image at higher resolution~(Fig.~\ref{fig::tablet_sizing}c--d), in which case the boundary of the physical tablet are highlighted to help visually identify the drawing area (Fig.~\ref{fig::tablet_sizing}d -- e).

\prepara
\noindent{\bf Cutting--Planes and Cross--Dimension Contouring --}
We found that participants enjoyed the capabilities of viewing and contouring in 3D, yet they still wanted to preserve the capability of contouring on the 2D surface, due to its merits of precision and ease of observing detailed structures. 
Additionally, 2D contouring leverages existing skills of 2D medical images interpretation~\mbox{(\mbox{[N3]}, \mbox{[D1]})}.
Therefore, while designing C4, we proposed a cross--dimension contour workflow, by combining fast and direct contouring in 3D and precise 2D contouring on a tracked 2D tablet.
This process includes three aspects:
{\bf(a)}~Upon finding the target, oncologists first perform contouring \textit{directly} in the 3D volume using a 3D brush~(as the design in \mbox{[I2]}, see Sec.~\ref{sec::prelim::procedure} and Fig.~\ref{fig::drawing}e--f).
The brush--based contour would result in a 3D colored volume.
{\bf(b)} Oncologists would then select a cutting--plane with one of three orientations (Fig.~\ref{fig::example_interface}e).
We did not include an arbitrary orientation, as understanding cross--sectional slices with arbitrary orientation is challenging for domain experts~\mbox{[D3]}.
The contours on each slices were generated based on the selected slicer orientation~(Fig.~\ref{fig::drawing}g).
{\bf(c)} Oncologists would finally inspect and revise contours mirrored on the tablet (Fig.~\ref{fig::drawing}a -- d). 
The inter--slice interpolation will be applied as needed to reduce the lead time (see Sec.~\ref{sec::method::implementations}).

\begin{figure}[h!]
    \centering
    \vspace{-0.05in}
    \includegraphics[width=\linewidth]{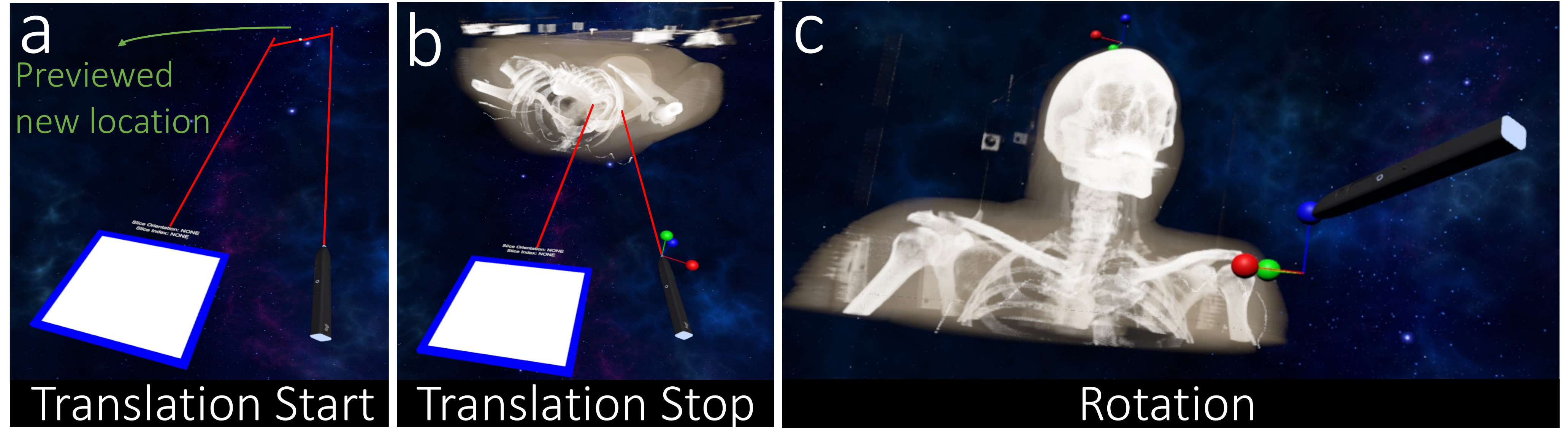}
    \vspace{-0.25in}
    \caption{Transforming 3D medical structure. (a) Two rays from the VR pen and tablet are initiated after translation start; (b) after the translation stop, the 3D medical structures are anchored to the new place; (c) during rotation mode, the quarternion of the near--hand anchor is projected onto the 3D medical structure.}
    \vspace{-0.15in}
    \label{fig::transformation}
\end{figure}

\prepara
\noindent{\bf Transforming 3D Medical Structures --}
During C3 and C4, participants were required to see the 3D medical structures while contouring, as well as to draw directly inside the 3D volume. 
It is therefore critical to design efficient ways for manipulating 3D medical structures inside VR.
To address this, we revisited and revised the design techniques of workspace translation and scaling in \mbox{SymbiosisSketch~\cite{Arora2018}}, which was evaluated among six design professionals.
After initiating the transforming mode, two rays are used to determine the new position where the 3D volume will be placed.
The new position is decided by the midpoint of the line segment whose length is the shortest between two rays~(Fig.~\ref{fig::transformation}a).
To avoid visual clutter, a 3D cursor is used to indicate the center of the newly placed 3D volume.
A second touchpad click allows the 3D volume to be moved into the target place~(Fig.~\ref{fig::transformation}b).
Oncologists can rotate the volume by clicking--and--dragging the near--hand anchor (Fig.~\ref{fig::transformation}c). 
The quaternion of the near--hand anchor is then mapped to the 3D medical structure.

\prepara
\noindent{\bf Support of Coarse-- and Fine--grained Tuning of Parameters --}
Finding optimal parameters has been identified as a critical factor that affects contour delineations~(\eg~locating the slices, changing the tablet and brush size, see [D7], [D8]). 
For coarsely tuning parameters, oncologists can use the scrollbar shown in front of them.
They can also click on the pen's touchpad for fine--grained parameter adjustments.
Although we enabled both tuning methods for all parameters during \mbox{[I2]}, oncologists suggested only adding fine--grained tuning to the {\it slice selections} due to its frequent uses during contouring.

\prepara
\noindent{\bf Spatially Distributed Control Panels --}
While only retaining essential features of general computer--supported contouring applications, we found that it was challenging for participants to promptly locate target UI elements if aggregated onto only one panel~\mbox{[D2]}. 
Therefore, we decided to dispatch UI elements onto three sub--control panels for C2 -- C4. 
Each panel has elements with similar features: {\it tablet and volume control}, {\it stylus control}, and {\it interpolation control}.
Through this design, oncologists are able to re--organize the UI layout and/or hide specific panels as needed.
Participants were satisfied with these design enhancement during \mbox{[I4]}.

%% file: 05-methods.tex
\setlength{\abovedisplayskip}{0pt} 
\setlength{\abovedisplayshortskip}{0pt}

\section{Study Methods}\label{sec::method}
\subsection{Participants}\label{sec::method::participants}
\begin{figure}[b]
    \centering
    \vspace{-0.20in}
    \includegraphics[width=\linewidth]{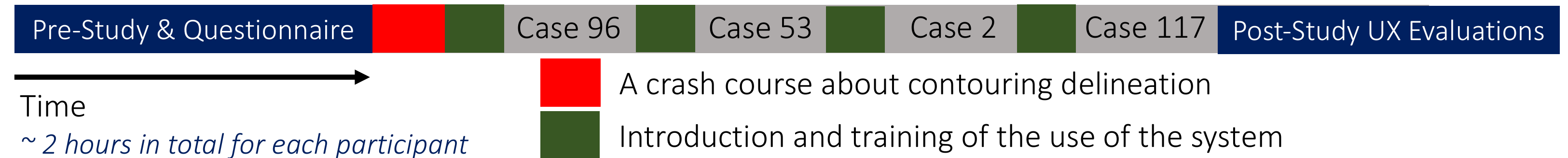}
    \vspace{-0.30in}
    \caption{Horizontal view of the study timeline.}
    \label{fig::timeline}
\end{figure}

We recruited $8$~participants~($2$~females and $6$~males, age: $M$ = $25.6$, $SD$ = $1.2$) who have taken human anatomy related curricula, but who {\it do not} yet had formal training on 2D medical image interpretation.
%
%
Our rationale behind this decision is based on two factors; 
\First, since we are evaluating the affordances of a specific modality, extensive past learning experience of 2D medical image interpretation could affect UX evaluations and system preferences;
\Second, participants taking human anatomy classes are able to identify and contour basic structures, even though clinical contouring is performed by senior residents and attending physicians.
To accommodate participants' anatomy knowledge, we intentionally selected fundamental structures with domain experts and made sure that the task was feasible for any students who have taken anatomy classes.
During our pre-study phase (see Fig.~\ref{fig::timeline}), participants overall {\it agreed} on a $5$-point Likert scale that they were able to identify key structures from 3D anatomy~($MED$ = $4$), yet {\it disagreed} that they could interpret 2D medical images~($MED$ = $2.5$).
Any participant that was involved with our design work (Sec.~\ref{sec::prelim}) was excluded from this evaluation.

\begin{figure}[t]
    \centering
    \includegraphics[width=\linewidth]{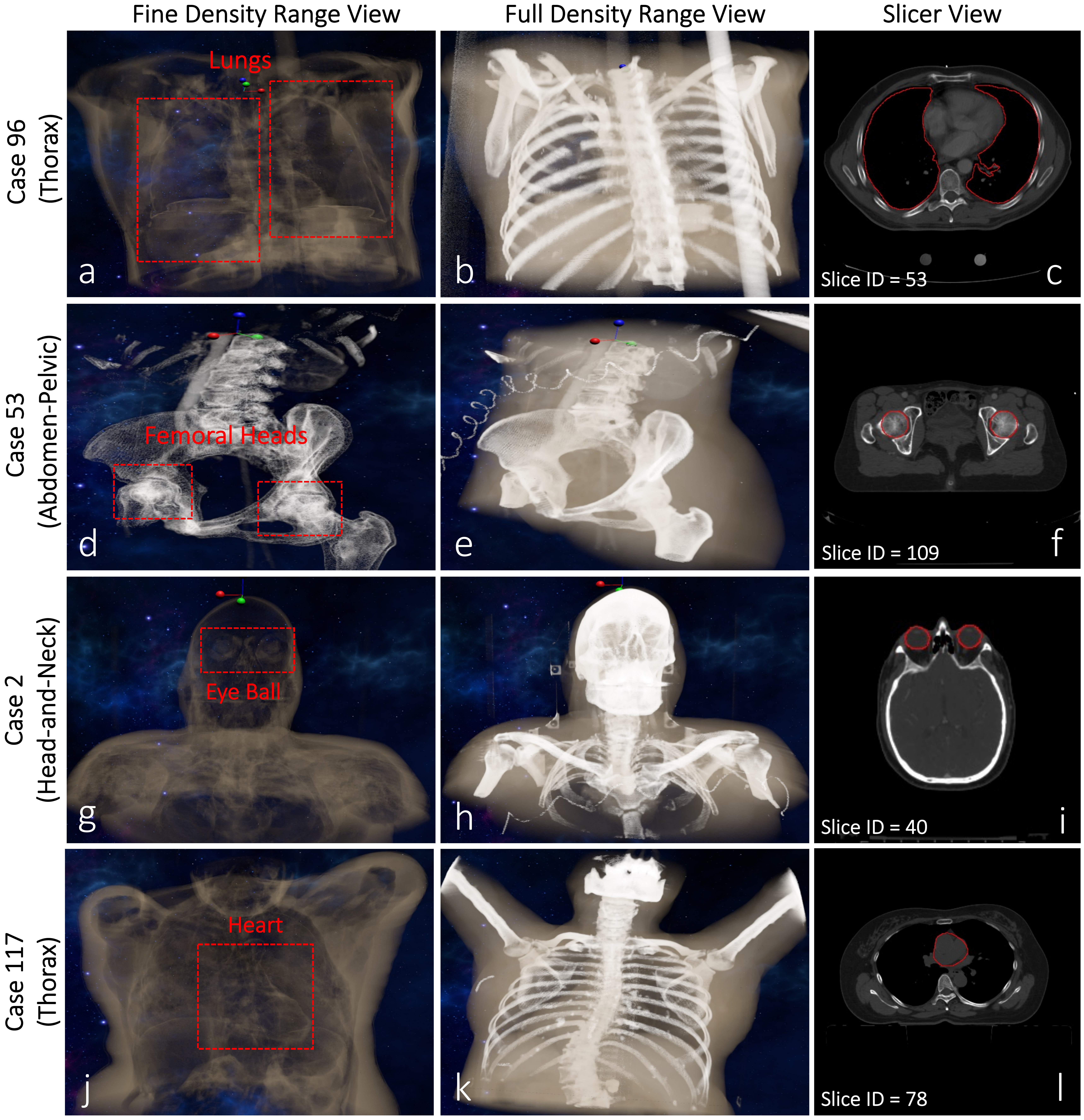}
    \vspace{-0.30in}
    \caption{Contour tasks: (a -- c) lungs from the thorax structure; (d -- f) femoral heads from the abdomen--pelvic structure; (g -- i) eye balls from the head--neck structure; (j -- l) heart from the thorax structure. Red contours in c, f, i, and l indicate expert contours. Displayed data are extracted from \href{https://econtour.org}{eContour}~\cite{eContour}, cases \href{http://econtour.org/cases/96}{$96$}, \href{http://econtour.org/cases/53}{$53$}, \href{http://econtour.org/cases/2}{$2$}, and \href{http://econtour.org/cases/117}{$117$}.}
    \vspace{-0.25in}
    \label{fig::rendering}
\end{figure}

\subsection{Contouring Tasks}\label{sec::method::tasks}
We decided to use the dataset from \href{https://econtour.org}{eContour}~\cite{Sherer2019}, which contains $119$~anonymized cases with expert contouring guidance.
Since investigating 3D rendering is out of our scope, we decided on four structures from four different patients with oncology educators that could be well visualized with our rendering pipeline~(see Sec.~\ref{sec::sys::rendering}).
%
%
Fig.~\ref{fig::rendering} demonstrates the 3D views of the selected tasks (with fine and full density ranges), along with one example transverse cutting--plane, as well as an expert contour (noted by yhe red outlines).

\subsection{Procedures}\label{sec::method::procedure}
We structured our within--subject study into three phases~(see Fig.~\ref{fig::timeline}).

\prepara
\noindent{\bf Phase 1: Pre--Study Orientations \& Questionnaire --}
Participants were first asked to completed a pre--study questionnaire, aiming to collect demographic information, VR experience, and the skills for identify key structures in 3D anatomy and from 2D scans. 
Participants were then introduced to the concept of contouring and inter--slice interpolation.

\prepara
\noindent{\bf Phase 2: VR Data Collection --}
We aim to evaluate participants' contouring performance of four conditions (see~Fig.~\ref{fig::design-space-explorations}).
To avoid the confounding factors of learning experience, a Latin square design~\cite{Dekking2005} was used to counterbalance the order of conditions.
This means that the contouring task (see Fig.~\ref{fig::rendering}) performed by each participants using a specific condition would be varied and balanced across participants.
Participants were given opportunities before each session to familiarize themselves with the system.

\prepara
\noindent{\bf Phase 3: Post--Study UX Evaluations --}
Participants were finally asked to fill out a questionnaire, and think-aloud while recording their responses.
System Usability Scale~(SUS)~\cite{Brooke1996sus} and NASA TLX~\cite{Hart1988} were used to evaluate their perceived usability and task load.
Participants were also asked to rate the usability of using all four conditions as tools to {\it learn} contouring, on a $5$--point Likert scale.
Brief semi--structured interviews were also conducted focusing on the merits and setbacks of seeing and directly drawing inside 3D medical structures.

\begin{figure*}[t]
    \centering
    \includegraphics[width=\linewidth]{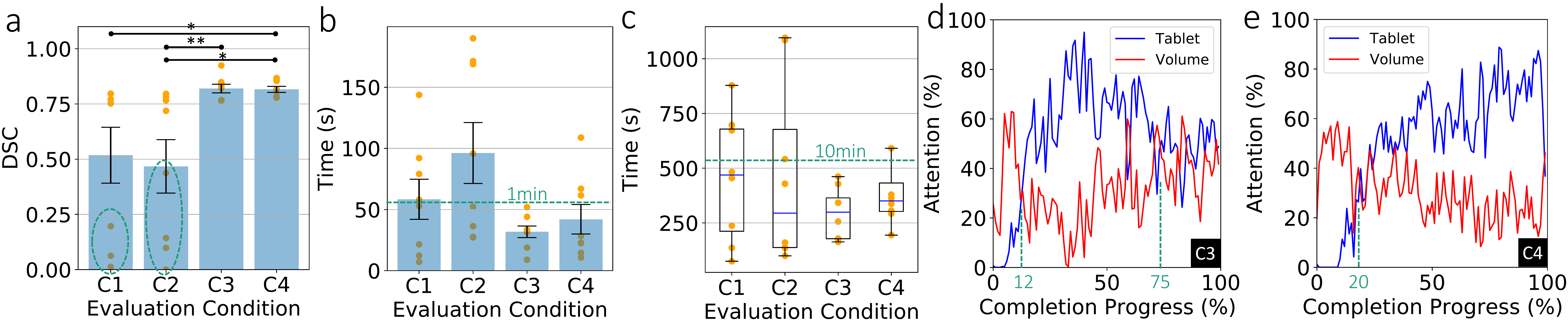}
    \vspace{-0.30in}
    \caption{Quantitative results~(\mbox{$* = p < .05$}, \mbox{$** = p < .01$}, \mbox{$*** = p < .001$}) for DSC (a), time for initial explorations (b), and overall TCT (c) measured on four conditions; standard error is used to visualize the error bar and the overlaid yellow scatters indicate individual measurements. Panels (d) and (e) show the average of attention ($\%$) that participants allocated to tablet and 3D volume while contouring using C3 and C4 respectively.}
    \vspace{-0.25in}
    \label{fig::quant_results}
\end{figure*}

\subsection{Quantitative Measures}\label{sec::measures}
%

\noindent{\bf Consensus Measurement --}
To evaluate the quality of participants' contour, we compute the Dice Similarity Coefficient (DSC)~\mbox{\cite{Zijdenbos1994}} between participants' and expert's contours. 
%
%
Equation~(\ref{eqn::dsc}) shows the computations of DSC, where $X_i$ and $Y_i$ refer to the contour masks of expert's and user's contour at $i^{th}$ slice.

\vspace{-0.1in}
\begin{equation}
    DSC = \frac{2 \sum_{i = 0}^{n} |X_i \cap  Y_i|}{\sum_{i = 0}^{n} |X_i| + \sum_{i = 0}^{n} |Y_i|}
    \label{eqn::dsc}
\end{equation}

\prepara
\noindent{\bf Temporal Domain Evaluations --}
We measure the time for initial explorations and overall Task--Completion--Time~(TCT).
The {\it time for initial exploration} refers to the duration \after the spatial UI being anchored and \before the first stroke being drawn.
During this step, participants need to identify and localize the majority of the target.
On the other hand, the {\it overall TCT} refers to the duration that participants take for completing the entire contouring procedures.

\prepara
\noindent{\bf Gaze Analysis --}
We analyzed the gaze direction to analyze {\it where did participant focus on during a particular fixation instant?}).
We first filter the frames with the gaze angular speed $\ge 150$$^\circ$$/$s, which are usually considered as saccade~\cite{Britannica1987}.
We then compute the \% of attention that participants allocated during each sliding window on {\it tablet}~(2D) and {\it volume}~(3D), quantified by the \% of frames when gaze collides with tablets and volume, respectively.
Since the overall TCT varies among individuals, we used the \% of the completion progress as the independent variable.
For simplicity, $1$\% of overall TCT is used as the width of sliding window, and the $\%$ of attention was computed upon every unit increment of task completion progress. 
Gaze analysis are only used to analyze C3 and C4, as only 2D interfaces are included in C1 and C2.

\subsection{Implementation}\label{sec::method::implementations}
We implemented all testbeds on HTC VIVE Pro Eye~\cite{HTCVivePro} using Unity~(v2020.3.6f1) and SteamVR~(v1.21.2).
Tobii XR SDK (v3.0.1.179)~\cite{tobiixr} was used to track gaze direction.
The rendering ran on a gaming laptop ($32$GB RAM, $12$-core CPU, and a RTX~$3080$ GPU).
Inter--slice interpolation was also integrated~\cite{Albu2008, Zukic2016nd}.
%
For C1, we used a keyboard~\cite{hi_keyboard} and mouse~\cite{logitech_mouse} as the input device, to emulate contouring on today's 2D display.
For C2 -- C4, we used a Logitech VR Pen~\cite{LogitechVRInk} and a tracked tablet as the input device (see Sec.~\ref{sec::system::interaction}).
we used three SteamVR Base Station (v2.0)~\cite{ValveBaseStation} for tracking purposes.
As investigations of tracking performance is out of our scope, we have adjusted the placements of base station and workspace and ensure that the pen and tablet are tracked during at least $99\%$ of the frames (see Appendix~\ref{sec::app::setup}).
We implemented a vertex and fragment shader based on \cite{unityvolumetricrendering}, for rendering cutting--planes and 3D volumes.
A compute shader was implemented for coloring each pixel/voxel.
For C3 and C4, participants' drawn contours would be mirrored on 3D medical structure while drawing on tablet, and vice versa.

%% file: 06-results.tex
\section{Results}\label{sec::results}
\subsection{Quantitative Measurements}\label{sec::results::quant}
Figure~\ref{fig::quant_results} shows quantitative results.
Repeated Measures Analysis of Variance (RM--ANOVA) were used to analyze various measures ($\alpha$ = $.05$).
We used Aligned Rank Transformation~(ART)~\cite{Wobbrock2011} for those that failed to pass the normality check, and ART procedures for multi--factor contrast test for post--hoc test, with Bonferroni corrections~\cite{artc2021}.
%
%

\prepara
\noindent{\bf Overall DSC --}
We first show that changing the evaluation condition (C1, C2, C3, or C4) might change the overall Dice Similarity Coefficient (DSC)~(\mbox{$F_{3, 28}=5.79$}, \mbox{$p = .005$}, \mbox{$\eta^2=0.45$}).
The subsequent post--hoc test demonstrates a statistical significance between C1 and C4~($p = .04$), C2 and C4~($p = .04$), as well as C2 and C3~($p = .02$).
By looking at the overlaid scatter plot in Fig.~\ref{fig::quant_results}a, we observed that three and four participants failed to locate the target structure in C1 and C2 (\textit{without} 3D visualizations), leading to $<50$\% of DSC~(\mbox{$M_1 = 51.80$\%}, \mbox{$M_2 = 46.76$\%}), and higher standard deviations compared to their counterpart (\mbox{$SD_1 = 13.65$\%}, \mbox{$SD_2 = 12.10$\%}~\vs~\mbox{$SD_3 = 1.93$\%}, \mbox{$SD_4 = 1.37$\%}). 
All participants could successfully identify the majority of the targets with C3 and C4 (\textit{with} 3D visualization), leading to $>75$\% of DSC~(\mbox{$M_3 = 81.98$\%}, \mbox{$M_4 = 81.61$\%}).
It is worth mentioning that the DSC measured from our study should not be considered as a reference for clinical purposes~(\eg~usually $>80$\% of DSCs were measured in contouring of cervical cancer~\cite{Viswanathan2014}), as all participants are pre--med or junior MD students without practical contouring experience.
Additionally, tracking inaccuracies could also hinder the consensus measurements.

\prepara\noindent
{\bf Time for Initial Exploration --}
RM-ANOVA (\mbox{$F_{3, 28}=3.03$}, \mbox{$p = .04$}, \mbox{$\eta^2=0.25$}) shows a marginal $p$--value, but large effect size, implying that our results show an effect, but are likely limited by the small sample size.
No statistical significance was detected during post--hoc test.
All participants located the target using C3 within $1$~min.

\prepara\noindent
{\bf Overall TCT --}
No statistical significance was detected when it comes to overall Task-Completion-Time (TCT). To illustrate our results we therefore visualize the boxplot shown in Fig.\ref{fig::quant_results}c.
Results indicate all participants were able to finish the instructed contour tasks within $10$~min with C3.
While it seems that the capability of direct contouring in 3D could accelerate the overall task by minimizing the procedures of delineating outlines, more time was spent for refining the outlines, compensating for the merits brought by the novel affordance given by direct 3D drawing.

\prepara
\noindent
{\bf Gaze Analysis --}
Fig.~\ref{fig::quant_results}d--e approximate the average of \% of attention that participants allocated on tablet~(2D) and volume~(3D) while progressing towards task completion.
With C3, participants heavily relied on the 3D volume for initial explorations~($\sim$ first $12$\% of progress).
When task completion progress reached more than $75$\%, both tablet and 3D volume were used for finalizing the contouring.
With C4, participants mainly used the 3D volume during initial explorations and 3D contours~($\sim$ first $20$\%).

\begin{figure*}[t]
    \centering
    \includegraphics[width=\linewidth]{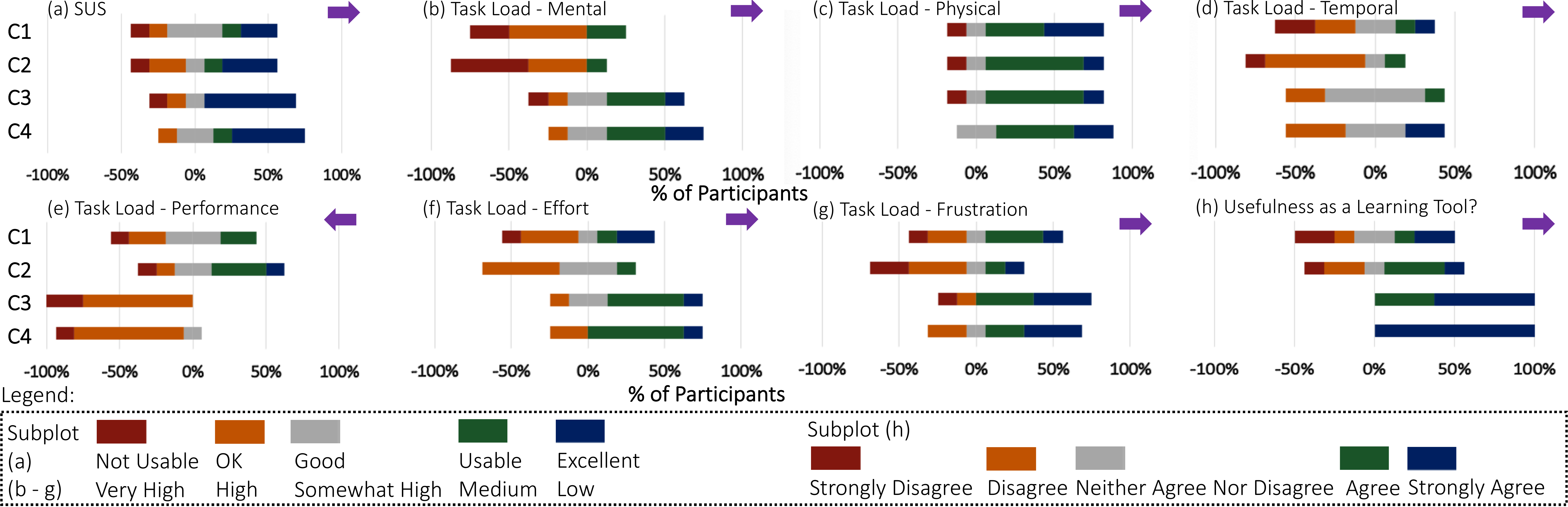}
    \vspace{-0.30in}
    \caption{Responses to post-study surveys. Purple arrow (top right) indicates the direction where a {\it good} system should be designed toward.}
    \vspace{-0.25in}
    \label{fig::ux}
\end{figure*}

\subsection{Post--Study UX Evaluations}\label{sec::results::ux}
We report here on participants' responses to SUS (Fig.~\ref{fig::ux}a), task load (Fig.~\ref{fig::ux}b -- g), and level of agreement of \textit{usefulness as a learning tool} (Fig.~\ref{fig::ux}h).
We used Sauro's~\cite{Sauro2018} and Hart~\etal's~\cite{Hart1988} recommendations to interpret SUS and task load scores respectively.
Overall, participants rated C3 ({\it with} 3D visualization, {\it without} 3D contouring) as the most usable prototype (\mbox{$MED_3$ = $87.5$} \vs~\mbox{$MED_4$ = $78.8$}, \mbox{$MED_2$ = $72.5$}, \mbox{$MED_1$ = $71.3$}).
``Excellent'' is considered for \mbox{SUS $\geq 80.8$} ~\cite{Brooke1996sus, Sauro2018}.
Four takeaways are generated through thematic analysis:

\prepara\noindent
{\bf Seeing 3D Medical Structures --}
All participants appreciated the merits of seeing the 3D structure inside VR, and think {\it ``the 3D was very useful''} [P5].
With the 3D visualization (C3 and C4), more participants reported less mental load (Fig.~\ref{fig::ux}b) and higher performance (Fig.~\ref{fig::ux}e).
First, participants felt that 3D visualizations are useful for understanding the structure ({\it ``when I just look at the slides, I'm not sure if that's the target or not, but if I see it in 3D, I'm more confident.''} [P3]), reduce mental load ({\it``It helps to have a bigger picture, and it takes off that cognitive load to actually imagine the 3D representation, which is a very high load.''} [P2]), and locating targets [P5].
%
%
%
P7 pointed out that the 3D visualization addresses one of the hardest challenges in anatomy class: {\it ``I think that the hardest part of anatomy is trying to look at pictures on a 2D surface, and then make them 3D, which is like exactly what this is accomplishing''} [P7].
P3 and P7 emphasized that the 3D medical structure helped them tremendously on deciding the boundary of the target structures (\eg~heart and lungs).
%
Some participants appreciated the transparent volumetric rendering: {\it``we use an app called Complete Anatomy, but I like [the transparent rendering] a lot better. You can see more detail and get depth. It's really hard to get depth for non--transparent rendering.''} [P7].
Second, with 3D structures, participants could better leverage the inter--slice interpolation by deciding the slice position where the rapid structure changes occurred 
(\eg~{\it ``it will help in locating the cutting-plane.''} [P5], {\it ``you can see the plane that you put in your virtual cadaver, and it helps feel where the rapid changes occur.''} [P7]).

\prepara\noindent
{\bf Drawing Contours in 3D --}
Participants appreciated the capabilities of direct drawing in 3D while attempting to locate the target (\eg~{\it``A huge part of the work gets done way faster''} [P2], {\it ``I like the mapping that map my annotation in 3D to 2D''} [P6]).
Some participants also appreciated the precision offered by the VR pen ({\it``[the pen] is intuitive for precise hand control''} [P3]).
However, we also observed participants expressing slightly higher frustration by comparing C3 and C4~($MED$: $12.5$\% \vs~$20$\%);
potential reasons include difficulties of drawing precisely in 3D 
(\eg~{\it ``It was very hard for me to be precise when drawing in 3D, I didn't see where I was starting''} [P2]), 
lack of capabilities for inspecting inside 3D volumes 
(\eg~{\it ``You cannot look inside the body. When you look inside the body, you cannot see very clearly where you are. [...] I think a better thing is as your pen is approaching, in the 3D model, you're looking at a slice of the 3D model, so that you know exactly where you are''} [P3]),
and small size (\eg~{\it ``if you could blow it up more, to be able to get finer precision, that would make it a lot easier''} [P7]).

\prepara\noindent{\bf Learning Cost --}
Echoing \mbox{[N2]} and \mbox{[N3]}, some participants also mentioned that {\it``medical school students mostly never used VR''} [P3].
While trained to use the system before each session, all participants found it not hard to learn and get familiar with system operations (\eg~{\it``a week for learning probably would be sufficient''} [P8]).
This echos the fact that seven participants {\it agree} or {\it strongly agree} with the statement that {\it ``I would imagine that most people would learn to use this system very quickly''} in SUS responses.

\prepara\noindent{\bf Contouring Education --}
%
All participants {\it agreed} or {\it strongly agreed} that seeing and drawing in 3D is extremely helpful for learning contouring (\eg~{\it``if you could put this into like, even our anatomy labs, that'd be amazing''} [P7]).
While comparing the experience of using C3 and C4, P1 outlined additional setbacks of learning anatomy with cadavers, including the unpleasant smell of formaldehyde, the mental stress triggered by a real body, and the reduced accessibility and resulting practicing opportunities.
%
P7 appreciated the transparent rendering while comparing it with \emph{Complete Anatomy}~\cite{complete_anatomy_2020}, a mobile application designed for anatomy learning purpose, because of the capability of {\it ``seeing more details and getting depth information''}.

%% file: 07-discuss.tex
\section{Discussion and Future Work}\label{sec::discussion}

\subsection{Seeing and Contouring in 3D}\label{sec::discuss:::viz_contour_3D}
Our results show that there is great potential for radiation oncology and contouring when integrating 3D visualizations into existing contouring workflows through VR.
While our focus is on RT treatment planning, many findings are more general and could be transferred to other similar applications that require annotations on 3D volumes (\eg~annotations during surgical planning~\cite{ARTEMIS2021}).

Since some of our participants felt that it was less intuitive and overall harder to directly draw in 3D -- mostly because of lack of depth perception and fine grained control -- one natural follow up would be to {\it investigate the design of novel interaction techniques where participants can draw precisely in 3D in these contexts}.

Another direction that future work might consider is the inherent challenges raised by interacting with 3D medical structures.
While the metaphor of pen $+$ tablet could offer precise drawing, some participants believe that they are less intuitive when it comes to 3D object manipulations.
Future investigations of interactions with 3D volume might be needed to uncover more insights.

\subsection{Re--thinking the Contouring Training Curriculum}\label{sec::dicuss::curriculum}
%
%
Contouring is a challenging skill to master, and it is typically trained during residency programs.
While interpreting 2D scans is usually difficult, it is a necessary prerequisite for learning contouring.
However, Dmytriw~\etal~\cite{Dmytriw2015} surprisingly showed that $< 10$\% of students are \textit{very} or \textit{completely} confident in interpreting basic radiology images.
Although additional training during residency program might address such gaps, many residents might not have that opportunity, especially in less developed areas~\cite{Holt2001, Sura2017, Eansor2021}.
%
With \mbox{VRContour} we unveil an alternative solution -- bringing the contouring workflow into a VR--based 3D UI -- that could possibly help radiation oncologists (and others) to massively scale contouring education. 

Although participants recognized that getting familiar with VRContour is not challenging, we also found that half participants did not have extensive experience on using VR (rated as {``only tried it out once or twice before''} in our pre--study questionnaire), and exhibited less experience with exploring 3D user interfaces.
One future direction is to rethink and modernize today's oncology educations by fully integrating VR as part of future curricula.

\subsection{Limitations}\label{sec::discussion::limitation}
The limitation of our work can be summarized into three areas.
\First, our results are based on $8$ participants, and the small sample size together with the lack of background/expertise/preference diversity might affect some results.
While we only considered participants without extensive experience in interpreting 2D scans to minimize the confounding factors of learning experience, such participants will also not have prior contouring experience, limiting the application of our results to different levels of contouring proficiency.
\Second, since examining participants' anatomy knowledge was beyond our scope, our study was based on only four cases extracted from \href{https://econtour.org}{eContour}~\cite{Sherer2019} during \mbox{[I4]}, with the goal to have baseline scenarios that would be familiar for any students with basic anatomy training.
However, practical contouring usually involves tumors and finer--grained anatomy, which might require prior experience on oncology and pathology, which would have been challenging for current participants.
Future work might introduce more realistic tasks, and include residents and attending physicians as part of the evaluation.
\Finally, we intentionally excluded procedures for adjusting volumetric rendering parameters from our study.
Instead, we pre--tuned parameters and offered {\it full} and {\it fine} density range view for participants to inspect the target and overall view of the 3D volume. 
However, adjusting such parameters is an inevitable step in practical contouring, and it might be considered in future work.

%% file: 08-conclusion.tex
\section{Conclusion}\label{sec::conclusion}
We presented {\it VRContour}, and investigated how to bring radiation oncology contouring into VR.
We explored and defined three design spaces by considering the dimensionality for consuming and inputting information. 
In a within -- subject study \mbox{(n = $8$)}, we found that the inclusion of a 3D visualization could increase contour precision and reduce mental load, frustration, and overall effort, as well as introduce important benefits for learning and educational purposes.
We believe that our work will benefit future research in designing VR--based medical software tools for radiation oncology and beyond.

%% file: appendix.tex
\renewcommand{\thefigure}{A\arabic{figure}}
\setcounter{figure}{0}

\section{Experimental Setups}\label{sec::app::setup}
This section provides supplementary material in addition to Sec.~\ref{sec::method::implementations}, specifically in terms of describing our experimental workspace.

\begin{figure}[h]
    \centering
    \includegraphics[width=\linewidth]{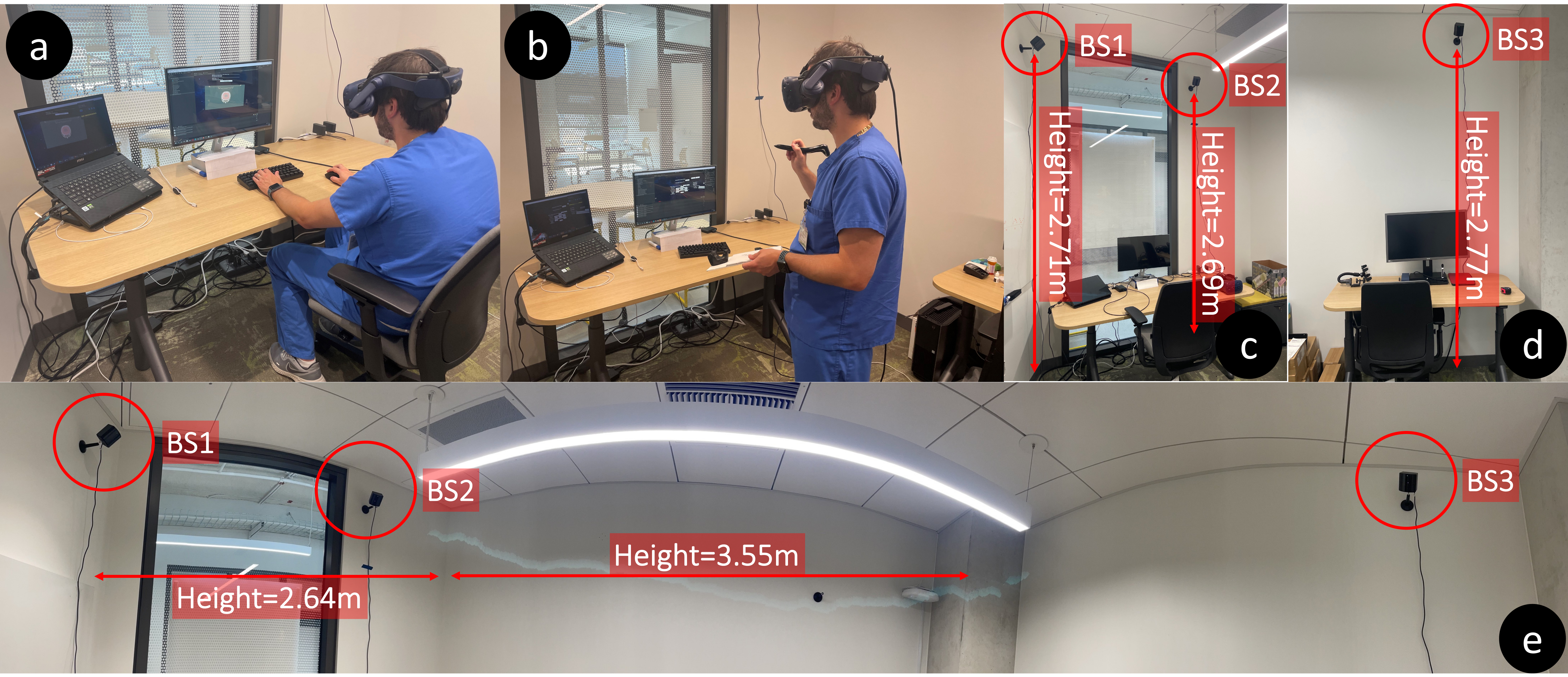}
    \vspace{-0.28in}
    \caption{Experimental setups. (a -- b) Participant could sit or stand while performing designated contouring tasks. (c) The base station mounted at the front of the room. (d) The base stations are mounted at the back of the room. (e) A panoramic view of the experimental office.}
    \vspace{-0.25in}
    \label{fig::experimental_room}
\end{figure}

While investigating tracking accuracy of pen and tablet is out of our scope, it is still important to maintain reasonable tracking performance for conducting all evaluation conditions. 
Fig.~\ref{fig::experimental_room} demonstrate the setups of the experimental office space, as well as the final placements of three SteamVR base stations~\cite{ValveBaseStation}.

